\documentclass[jkps,12pt,showpacs,showkeys]{revtex4}

\usepackage{graphicx}
\usepackage{amssymb,amsmath}

\begin{document}

\title[]{Growing Probability of Additional Offspring with a Beneficial Reversal Allele in the Decoupled Continuous-time Mutation-selection Model}

\author{Wonpyong \surname{Gill}}
\email{wpgill@pusan.ac.kr}
\affiliation{Department of Physics, Pusan National University, Busan 609-735}

\date[]{Received 2 February 2015}

\begin{abstract}

The growing probabilities of additional offspring with the beneficial reversal allele for various population sizes, $N$, sequence lengths, $L$, selective advantages, $s$, fitness parameters, $k$, and measuring parameters, $C$, were calculated in an asymmetric sharply-peaked landscape using the decoupled continuous-time mutation-selection model.
The growing probability in the stochastic region was inversely proportional to the measuring parameter when $C<1/Ns^*$, bent when $C\approx 1/Ns^*$ and became saturated when $C>1/Ns^*$, where $s^*$ is the effective selective advantage.
The saturated growing probability in the stochastic region was approximately the effective selective advantage when $C\gg 1/Ns^*$ and $s^*\ll 1$.
The present study suggests that the growing probability in the stochastic region in an asymmetric sharply-peaked landscape in the decoupled continuous-time mutation-selection model can be described using the theoretical formula for the growing probability in the Moran two-allele model.
The selective advantage ratio, which represents the ratio of the effective selective advantage to the selective advantage, does not depend on the population size, selective advantage, measuring parameter, and fitness parameter; instead the selective advantage ratio decreases with increasing the sequence length.

\end{abstract}


\maketitle

\section{INTRODUCTION}

Two mechanisms, random mutation and natural selection, are considered in the mutation-selection model to explain an evolutionary process. The mutation-selection model is generally classified into two types, the coupled model and the decoupled model, depending on the mutation mechanism considered. The coupled model considers mutations that are only allowed as replication errors during reproduction events. The decoupled model considers mutations that are allowed at any time during the life cycle, such as those due to the effects of radiation, free radicals or thermal fluctuations\cite{1,2,3,4,5}.
Hence, mutation and selection are assumed to be independent processes that occur in parallel.

From a biological point of view, the relative contributions of the underlying mutation mechanisms for the two versions of the mutation-selection model are controversial. A mutation mechanism described by the coupled version implies a relatively constant mutation rate per generation, whereas the mutation mechanism described by the decoupled version implies a relatively constant mutation rate in time.
Crow and Kimura\cite{1} and Akin\cite{2} studied the decoupled mutation-selection model.
Baake {\it et al.} reported that the decoupled mutation-selection model was equivalent to an Ising quantum spin and calculated the observable quantities in three fitness landscapes using the methods from statistical mechanics\cite{3,4,5}.
Recently, Saakian {\it et al.}\cite{6} and Park and Deem\cite{7} examined the decoupled continuous-time mutation-selection model using a functional integral representation.

Several studies have considered the evolutionary process transiting to a neighboring higher fitness peak through the low-fitness valley.
Iwasa et al. calculated the rate of stochastic tunneling in a large population when the intermediate mutant had a lower fitness than the wild type, but the second mutant was beneficial\cite{8}. Weissman et al. provided a complete description of the rate at which an asexual population traversed a specific path through a genotype space to a particular fitter genotype by neglecting backward mutations and genetic diversity\cite{9}. Biological examples of the low-fitness valley can be found in pathogens that escape the immune response of their hosts, such as Escherichia coli\cite{10,11}, Salmonella typhimurium\cite{12} and the human immunodeficiency virus\cite{13}. Another example might be the types of cancer that occur only after a series of mutations\cite{14}.

Previous studies calculated the crossing time and the growing probability in the coupled discrete-time mutation-selection model by switching on an asymmetric sharply-peaked landscape, such that the reversal allele of the optimal allele had higher fitness than the optimal allele, from an initial steady state in a sharply-peaked landscape\cite{15,16,17,18}.
The growing probability was defined as the probability that additional offspring with the reversal allele grow to a measuring criterion.
The present study calculated the growing probability in the decoupled continuous-time mutation-selection model, which was evaluated as the ratio of the first arrival time to the final arrival time. The first arrival time was defined as the time when the first additional offspring arrived at the reversal allele. The final arrival time was defined as the arrival time of a successful offspring with the reversal allele that could grow to a measuring criterion.

The growing probability of additional offspring with the reversal allele in the decoupled continuous-time mutation-selection model was calculated systematically using computer simulations for various population sizes, sequence lengths, selective advantages, fitness parameters, and measuring parameters, and was compared with the theoretical formula for the growing probability in the Moran two-allele model\cite{19,20}.
This paper is organized as follows: Section II briefly describes the decoupled continuous-time mutation-selection models and the computer simulation method.
Section III presents the computer simulation results for the growing probability in the decoupled continuous-time mutation-selection models. Section IV summarizes the results.

\section{Model and Method} \label{sec-2}

This study considered the decoupled continuous-time mutation-selection model in a self-reproducing haploid population of $N$ individuals, whose size remains constant over the generations. The multiple alleles, $S_i$, might be represented by a string of $L$ spins, which use two values, $+1$ and $-1$.
The time dependence of the number, $n_i$, of individuals with the allele, $S_i$, in the decoupled continuous-time mutation-selection model can be described by the following equation:
\begin{equation} \label{eq-1}
\dot{n}_i (t) = ( w_i - \bar{w} )\: n_i(t)\: + \: \sum_j Q_{ij} \: n_j (t) \ ,	
\end{equation}
where $w_i$ is the fitness of the allele $S_i$, and $\bar{w} (= {1 \over N} \sum_j w_j n_j(t) )$ is the mean fitness of the population.
In Eq. (\ref{eq-1}), the mutation matrix, $Q_{ij}$, which gives the mutation rate from $S_j$ to $S_i$, excludes double mutations\cite{4,5}:
\begin{equation} \label{eq-2}
Q_{ij}=\begin{cases}
	-L \mu \ , &\mathrm{if}\ i=j\ , \cr
	\mu \ , &\mathrm{if}\ d_{ij} = 1\ , \cr
	0 \ ,  &\mathrm{otherwise}\ ,
	\end{cases}
\end{equation}
where $\mu$ denotes the mutation rate per sequence element per unit time, and $d_{ij}$ denotes the Hamming distance between $S_i$ and $S_j$.

Consider a sharply-peaked landscape such that the optimal allele $S_0=(+1,+1, \cdots, +1)$ has a maximum fitness 1, and the remaining alleles have a lower fitness, $1-kL$.
The parameter $k$ is called the fitness parameter.
Equation (\ref{eq-1}) was solved using a computer simulation with the assumption that all individuals had the optimal allele initially. The population then approached a stochastic steady state, which is a quasispecies near the optimal allele in a sharply-peaked landscape. 
After arriving at a stochastic steady state, the sharply-peaked landscape was changed suddenly to the following asymmetric sharply-peaked landscape: 
\begin{equation} \label{eq-3}
w=\begin{cases}
	1 \ , &\mathrm{if} \ S=S_0 \ , \cr
	1+s \ ,  & \mathrm{if} \ S=S_0^\ast \ , \cr 
	1-kL \ ,  &\mathrm{otherwise}\ ,
	\end{cases}
\end{equation}
where the maximum fitness was assigned to the reversal allele, $S_0^\ast = (-1,\cdots, -1)$, and the fitness of the remaining alleles was unchanged. All other parameters were fixed in this procedure, and the parameter, $s$, in Eq. (\ref{eq-3}) is called the selective advantage of the reversal allele over the optimal allele. The reversal allele had higher fitness than the optimal allele, assuming that the selective advantage, $s$, is a positive value. The asymmetric sharply-peaked landscape had two fitness peaks that were separated by a low-fitness valley with a tunable depth and width.

The numbers, $n_i$, were grouped into $L+1$ distinct classes according to their Hamming distances from the optimal allele, $S_0$, to improve the speed of the computer simulation\cite{21}.
The number, $N_l(=\sum_{\{i\}_{d=l}} n_i)$, of individuals in class $l$ was defined by the sum of all numbers $n_i$ with a Hamming distance, $d=l$, from the optimal allele.
The kinetic equation, Eq. (\ref{eq-1}), for the decoupled model can be expressed using the newly defined $N_l$ as follows: 
\begin{equation} \label{eq-4}
\dot{N}_l (t)= (w_l - \bar{w})\: N_l(t)\: +\: \sum^L_{m=0} T_{lm}\: N_m (t) \ ,	
\end{equation}
\begin{equation} \label{eq-5}
T_{lm}=\begin{cases}
	-L \mu \ , &\mathrm{if}\ l=m\ , \cr
	(L-m) \mu \ , &\mathrm{if}\ l=m+1\ , \cr
	m \mu \ , &\mathrm{if}\ l=m-1\ , \cr
	0 \ ,  &\mathrm{otherwise}\ ,
	\end{cases}
\end{equation}
where $T_{lm}$ gives the mutation rate from class $m$ to class $l$.
The relative density, $X_l$, of class $l$ was then calculated by dividing the number, $N_l$, by the population size.

The computer simulations used the logistic branching process with the replacement of $\dot{N}_l (t)$ in Eq. (\ref{eq-4}) by $(N_l(t+dt)-N_l(t))/dt$\cite{19,22,23,24}.
The simulated population was evolved using the time steps of $dt=0.1$ generations in this study.
The mean fitness, $\bar{w}$, of the population was calculated at the beginning of each time step.
The branching process assumes that the population in the next time step is determined by the following two events: each individual in a class can mutate into any class with a certain probability, and each individual in a class can produce an offspring in the same class with a certain probability. More details of the simulation for a finite population can be found in Refs. 19 and 24.
The time was set to $t=0$ when the fitness landscape changed suddenly to an asymmetric sharply-peaked landscape. The increase in the relative density $\Delta X_d (t) (=X_d (t) - X_d (t=0) )$ was calculated. The computer simulation was completed when the increase in relative density with the reversal allele, $\Delta X_L (t)$, achieved the measuring criterion $\Delta X_L (t)=C$. The parameter, $C$, is called the measuring parameter.

The computer simulation recorded all arrival times, $t$, satisfying two conditions, $\Delta X_L(t-1) \le 0$ and $\Delta X_L(t) > 0$. The first arrival time, $t_a$, was defined as the earliest among the recorded arrival times, and the final arrival time was defined as the latest after the computer simulation is complete. Therefore, the final arrival time means the arrival time of a successful offspring with the reversal allele that can grow to a size, $NC$. The growing probability of additional offspring with the reversal allele was evaluated as the ratio of the first arrival time to the final arrival time, assuming that the additional offspring with the reversal allele are supplied at average time intervals of $t_a$ by stochastic fluctuations. This assumption holds well in the stochastic region, where the arrival of additional offspring with the reversal allele is a rare event with a quite long interval. Fortunately, the main consequence of this study was obtained from the simulation results for the growing probability in the stochastic region.

\section{Results} \label{sec-3}

\begin{figure}[h]
\includegraphics[width=10cm]{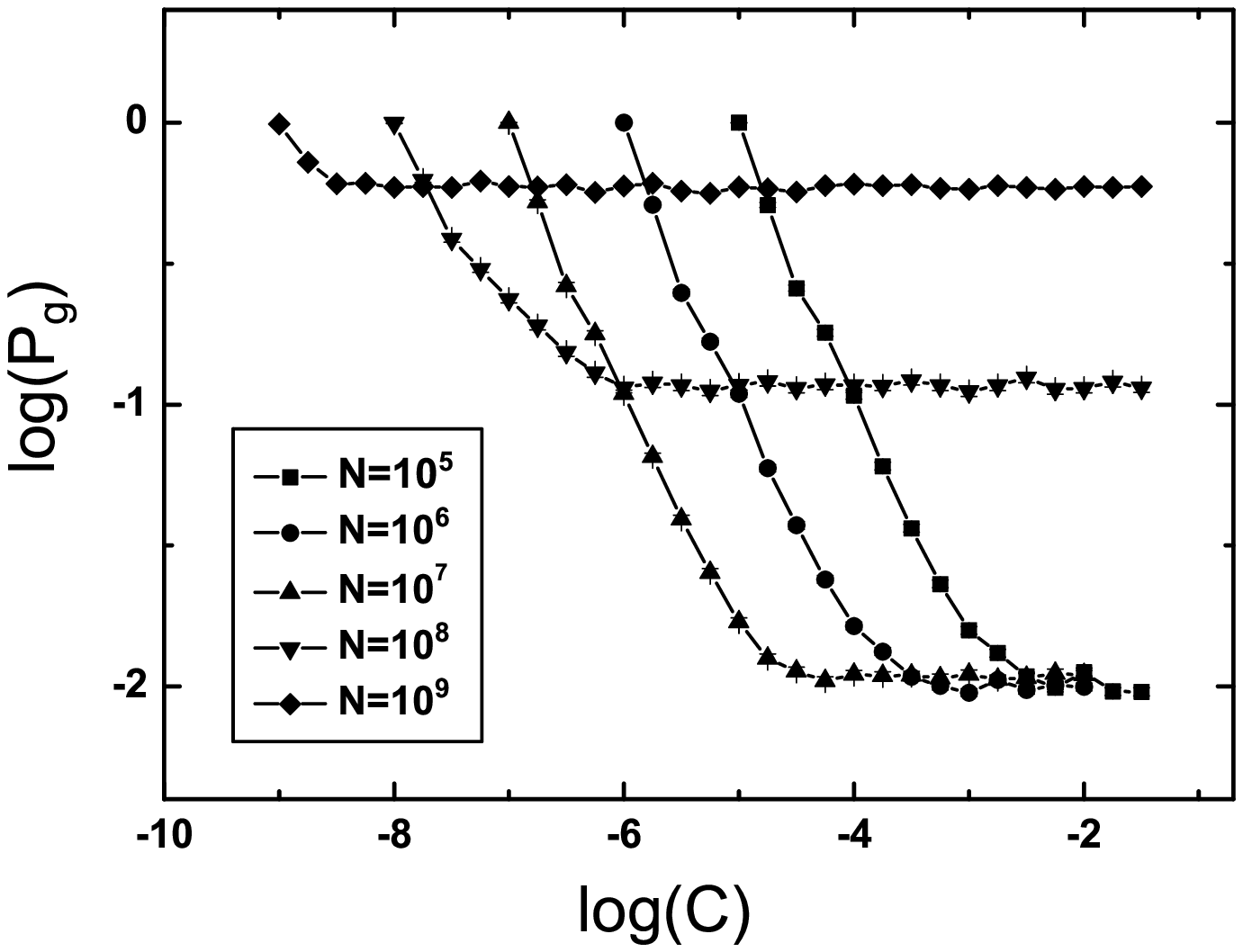}
\caption[0]{\label{fig-1} Logarithm of the growing probability, $\log(P_g)$, as a function of the logarithm of the measuring parameter, $\log(C)$, for $L=10$, $\mu =0.001$, $k=0.003$ and $s=10^{-4}$.
The simulation results for the growing probability in the decoupled continuous-time mutation-selection model are represented by the solid lines with symbols, which correspond, right to left, to the population sizes $N=10^5, 10^6, 10^7, 10^8$ and $10^9$.}
\end{figure}

The computer simulation results for the growing probability were obtained by averaging $M$ independent runs. The mean and the standard error were calculated from $M$ data values transformed into a logarithmic scale because the data showed a Gaussian distribution on a logarithmic scale.
The number of runs was set to $M=3000$ throughout this study.
The computer simulation results for an infinite population were calculated using the same method reported elsewhere\cite{24}.
Hereafter, $\Delta X_L^\infty (t)$ denotes the computer simulation result in an infinite population and $\Delta X_L(t)$ denotes the computer simulation result in a finite population.
Figure \ref{fig-1} shows the logarithm of the growing probability, $\log(P_g)$, as a function of the logarithm of the measuring parameter, $\log(C)$, for $L=10$, $\mu =0.001$, $k=0.003$ and $s=10^{-4}$.
In Fig. \ref{fig-1}, the simulation results for the growing probability in the decoupled continuous-time mutation-selection model are represented by the solid lines with symbols, which correspond, right to left, to the population sizes $N=10^5, 10^6, 10^7, 10^8$ and $10^9$.

A previous study showed that the boundary between the deterministic and stochastic regions in the decoupled continuous-time mutation-selection model could be determined using the critical population size, $N_c$, satisfying $N_c\Delta X^\infty_L(t=0)=1$\cite{24}.
The computer simulation for an infinite population showed that $\Delta X^\infty_L(t=0)\approx 4.73\times 10^{-9}$ and $N_c\approx 2.12\times 10^8$ for $L=10$, $\mu=0.001$ and $k=0.003$.
A previous study reported that the growing probability in the stochastic region in the coupled discrete-time mutation-selection model was inversely proportional to the measuring parameter when $C<1/Ns$, bent when $C\approx 1/Ns$, and saturated when $C>1/Ns$\cite{18}.
 
Figure \ref{fig-1} shows that the simulation results for the growing probability for $N=10^5, 10^6$ and $10^7$, which belong to the stochastic region because $N\ll N_c$, are inversely proportional to the measuring parameter when $C<1/Ns^*$, bent when $C\approx 1/Ns^*$, and saturated when $C>1/Ns^*$, where $s^*$ is the effective selective advantage.
The simulation suggested that the effective selective advantage, $s^*$, is a hundred times larger than the selective advantage, or $s^* \approx 100 s \approx 10^{-2}$.
Let the selective advantage ratio, $r$, be the ratio of the effective selective advantage to the selective advantage($r=s^*/s$).
The simulation results suggested $r \approx 100$ for the parameter region in Fig. \ref{fig-1}.
The simulation indicated that the number of additional offspring with the reversal allele, $n_L$, in the stochastic region drifts neutrally until $n_L$ reaches the size, $NC$, when  $n_L<1/s^*$, and that it grows deterministically when  $n_L>1/s^*$. 
Figure \ref{fig-1} also shows that the saturated growing probabilities, $P^*_g$, for $N=10^5, 10^6$ and $10^7$ become approximately the effective selective advantage, $s^* \approx 10^{-2}$, when $C\gg 1/Ns^*$ and $s^*\ll 1$.

\begin{figure}[h]
\includegraphics[width=10cm]{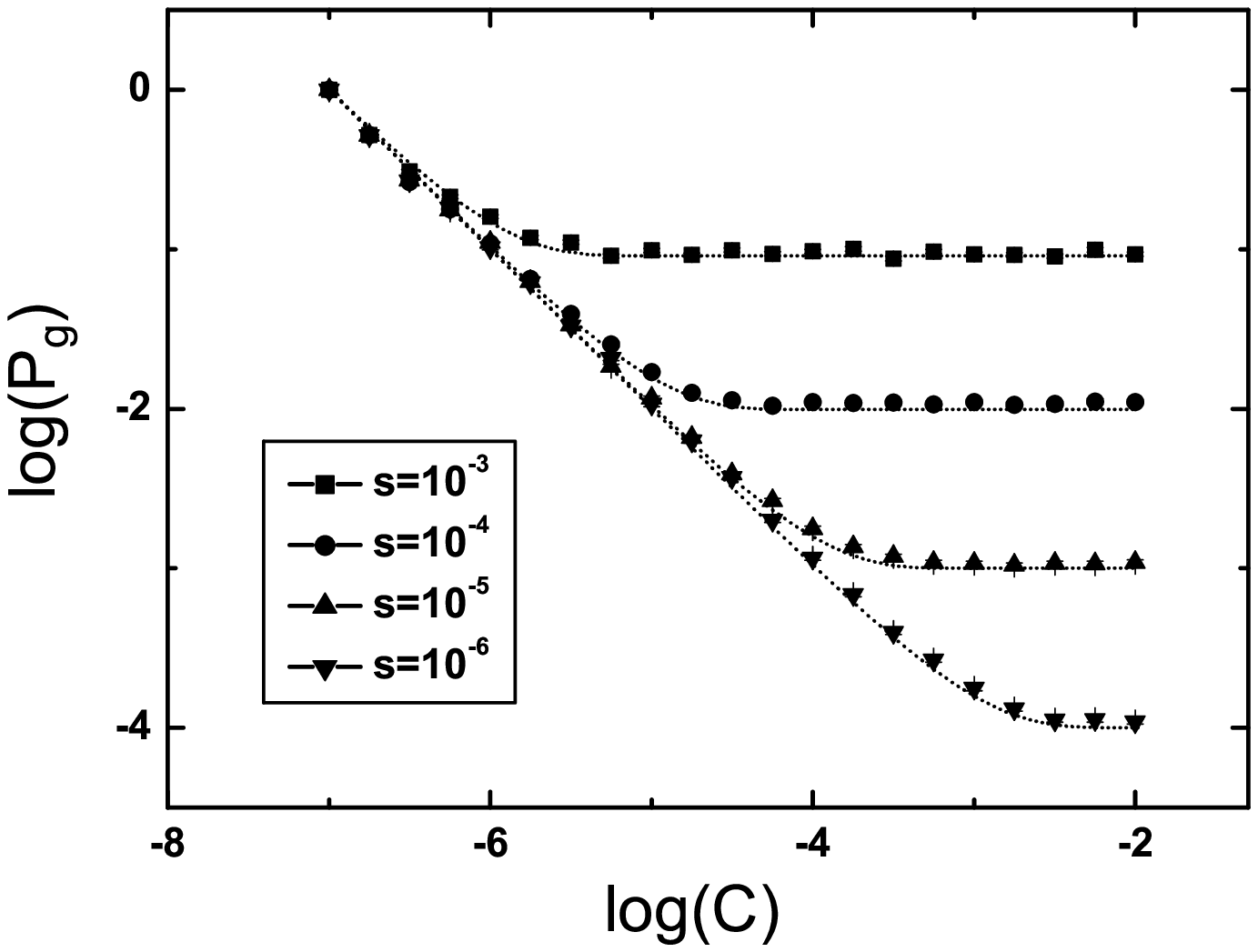}
\caption[0]{\label{fig-2} Logarithm of the growing probability, $\log(P_g)$, as a function of the logarithm of the measuring parameter, $\log(C)$, for $L=10$, $\mu =0.001$ and $k=0.003$.
The simulation results for the growing probability in a finite population with a size, $N=10^7$, in the decoupled continuous-time mutation-selection model are represented by the symbols, and the theoretical formula in Eq. (\ref{eq-6}) with $N=10^7$ and $s^*=100s$ is represented by the short-dotted lines, which correspond, top to bottom, to the selective advantages, $s=10^{-3}$, $10^{-4}$, $10^{-5}$ and $10^{-6}$.}
\end{figure}

Figure \ref{fig-2} shows the logarithm of the growing probability, $\log(P_g)$, as a function of the logarithm of the measuring parameter, $\log(C)$, for $L=10$, $\mu =0.001$ and $k=0.003$.
In Fig. \ref{fig-2}, the simulation results for the growing probability in a finite population with a size, $N=10^7$, in the decoupled continuous-time mutation-selection model are represented by the symbols, which correspond, top to bottom, to the selective advantages, $s=10^{-3}$, $10^{-4}$, $10^{-5}$ and $10^{-6}$.
The parameters in Fig. \ref{fig-2} belong to the stochastic region, $N\ll N_c$, because the critical population size, $N_c$, is independent of the selective advantage. Figure \ref{fig-2} clearly shows that the growing probabilities for various selective advantages are inversely proportional to the measuring parameter when $C<1/Ns^*$, bend when $C\approx 1/Ns^*$ and become saturated when $C>1/Ns^*$, where $s^* \approx 100 s$.
Figure \ref{fig-2} also shows that the saturated growing probabilities, $P^*_g$, for various selective advantages become approximately the effective selective advantage when $C\gg 1/Ns^*$ and $s^*\ll 1$.

The following theoretical formula for the growing probability, $P^t_g$, represents the probability for a beneficial allele to grow to a size, $NC$, in the Moran two-allele model without mutations, when the selective advantage between two alleles is assumed to be equal to the effective selective advantage, $s^*$\cite{19,20}:
\begin{eqnarray}
P_g^t &=&\ {s^* \: (1+s^*)^{NC-1} \over (1+s^*)^{NC} - 1} \ , \label{eq-6} \\
&\approx &\ \begin{cases}
	{s^* \over 1+s^*} \ , &\mathrm{if} \ s^*>1,\; s^*NC \gg 1 \ , \cr
	\ s^* \ ,  &\mathrm{if}\ s^*\ll 1,\; s^*NC \gg 1\ , \cr
	{1 \over NC} \ ,  &\mathrm{if}\ s^*NC \ll 1\ . \nonumber
	\end{cases}
\end{eqnarray}
In Fig. \ref{fig-2}, the theoretical formula in Eq. (\ref{eq-6}) with $N=10^7$ and $s^* =100s$ is represented by the short-dotted lines, which correspond, top to bottom, to the selective advantages $s=10^{-3}$, $10^{-4}$, $10^{-5}$ and $10^{-6}$. A comparison of the symbols and the short-dotted lines shows that the simulation results for the growing probability, $P_g$, can be fitted well using the theoretical formula for the growing probability, $P^t_g$, in the Moran two-allele model.
A previous study showed that the growing probability in the stochastic region in the coupled discrete-time mutation-selection model could be described using the theoretical formula for the growing probability in the Wright-Fisher two-allele model\cite{18}.
The present study suggests that the growing probability in the stochastic region in the decoupled continuous-time mutation-selection model can be described using the theoretical formula for the growing probability in the Moran two-allele model.

\begin{figure}[h]
\includegraphics[width=10cm]{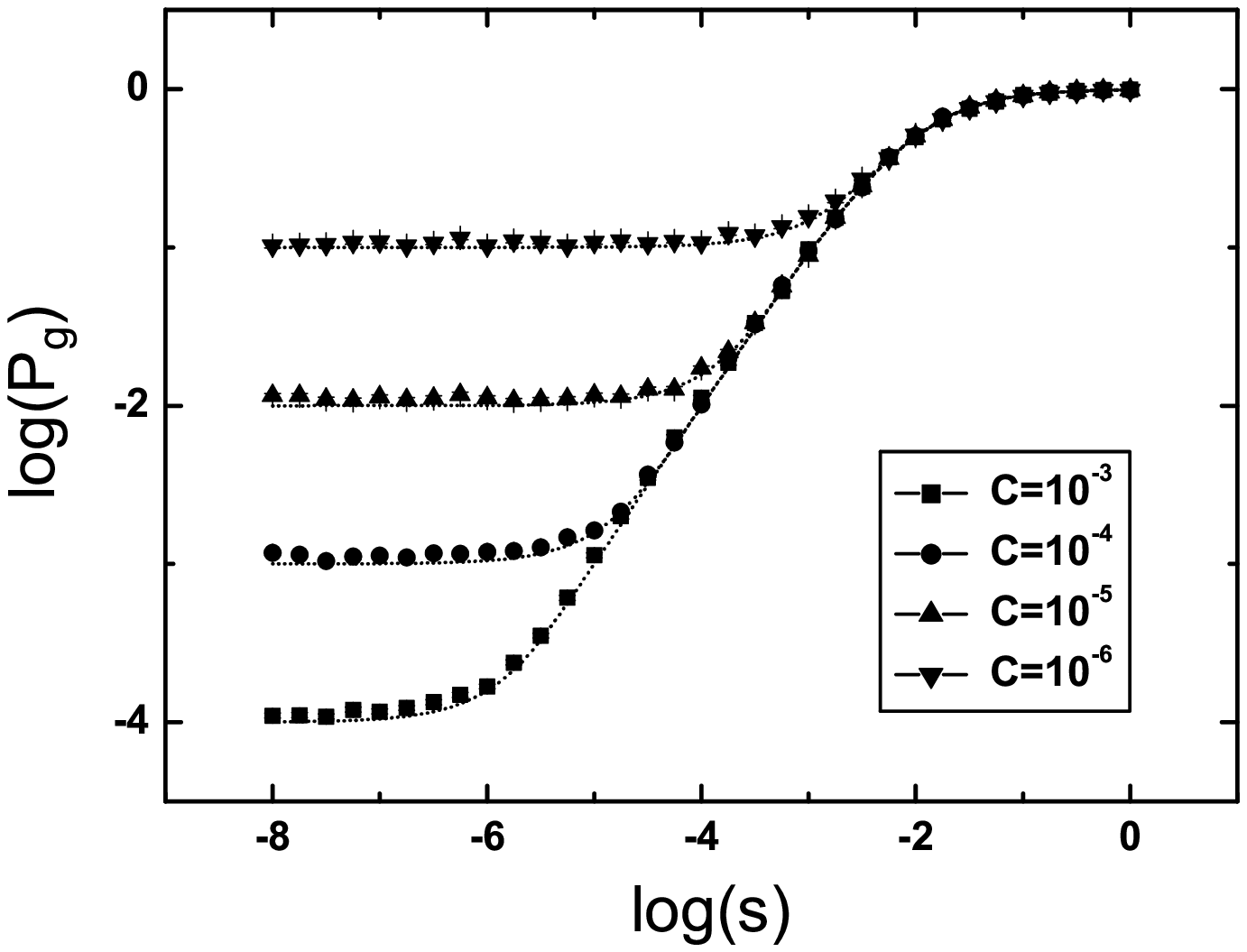}
\caption[0]{\label{fig-3} Logarithm of the growing probability, $\log(P_g)$, as a function of the logarithm of the selective advantage, $\log(s)$, for $L=10$, $\mu =0.001$ and $k=0.003$.
The simulation results for the growing probability in a finite population with a size, $N=10^7$, in the decoupled continuous-time mutation-selection model are represented by the symbols, and the theoretical formula in Eq. (\ref{eq-6}) with $N=10^7$ and $s^*=100s$ is represented by the short-dotted lines, which correspond, bottom to top, to the measuring parameters, $C=10^{-3}$, $10^{-4}$, $10^{-5}$ and $10^{-6}$.}
\end{figure}

Figure \ref{fig-3} shows the logarithm of the growing probability, $\log(P_g)$, as a function of the logarithm of the selective advantage, $\log(s)$, for $L=10$, $\mu =0.001$ and $k=0.003$. In Fig. \ref{fig-3}, the symbols represent the simulation results for the growing probability in the decoupled continuous-time mutation-selection model in a finite population with a size, $N=10^7$, for a range of measuring parameters, $C=10^{-3}$, $10^{-4}$, $10^{-5}$ and $10^{-6}$.
The parameters in Fig. \ref{fig-3} belong to the stochastic region, $N\ll N_c$, because the critical population size, $N_c$, is independent of the measuring parameter.
Figure \ref{fig-3} shows that the growing probabilities approach $1/NC$ when $s^*\ll 1/NC$, where $s^* \approx 100 s$.
In Fig. \ref{fig-3}, the theoretical formula in Eq. (\ref{eq-6}) with $N=10^7$ and $s^*=100s$ is represented by the short-dotted lines, which correspond, bottom to top, to the measuring parameters $C=10^{-3}$, $10^{-4}$, $10^{-5}$ and $10^{-6}$. A comparison of the symbols and the short-dotted lines shows that the simulation results for the growing probability, $P_g$, can be fitted well using the theoretical formula for the growing probability, $P^t_g$, in the Moran two-allele model.

\begin{figure}[h]
\includegraphics[width=10cm]{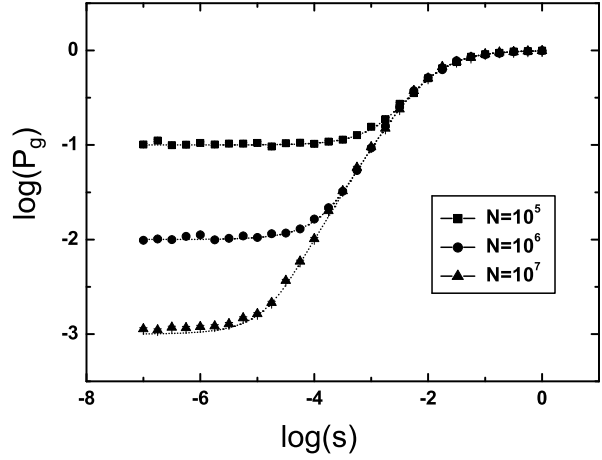}
\caption[0]{\label{fig-4} Logarithm of the growing probability, $\log(P_g)$, as a function of the logarithm of the selective advantage, $\log(s)$, for $L=10$, $\mu =0.001$, $k=0.003$ and $C=10^{-4}$.
The simulation results for the growing probability in the decoupled continuous-time mutation-selection model are represented by the symbols, and the theoretical formula in Eq. (\ref{eq-6}) with $C=10^{-4}$ and $s^*=100s$ is represented by the short-dotted lines, which correspond, top to bottom, to the population sizes, $N=10^5$, $10^6$ and $10^7$.}
\end{figure}

Figure \ref{fig-4} shows the logarithm of the growing probability, $\log(P_g)$, as a function of the logarithm of the selective advantage, $\log(s)$, for $L=10$, $\mu =0.001$, $k=0.003$, and $C=10^{-4}$. In Fig. \ref{fig-4}, the symbols represent the simulation results for the growing probability in the decoupled continuous-time mutation-selection model for a range of population sizes, $N=10^5$, $10^6$ and $10^7$.
The parameters in Fig. \ref{fig-4} belong to the stochastic region, $N\ll N_c$, as mentioned in the explanation of Fig. \ref{fig-1}.
Figure \ref{fig-4} shows that the growing probabilities approach $1/NC$ when $s^*\ll 1/NC$, where $s^* \approx 100 s$.
In Fig. \ref{fig-4}, the theoretical formula in Eq. (\ref{eq-6}) with $C=10^{-4}$ and $s^*=100s$ is represented by the short-dotted lines, which correspond, top to bottom, to the population sizes, $N=10^5$, $10^6$ and $10^7$.
The simulation results in Figs. \ref{fig-2}, \ref{fig-3} and \ref{fig-4} show that the growing probability, $P_g$, in the stochastic region in the decoupled continuous-time mutation-selection model for various population sizes, selective advantages, and measuring parameters can be fitted well using the theoretical formula for the growing probability, $P^t_g$, in the Moran two-allele model.

\begin{figure}[h]
\includegraphics[width=10cm]{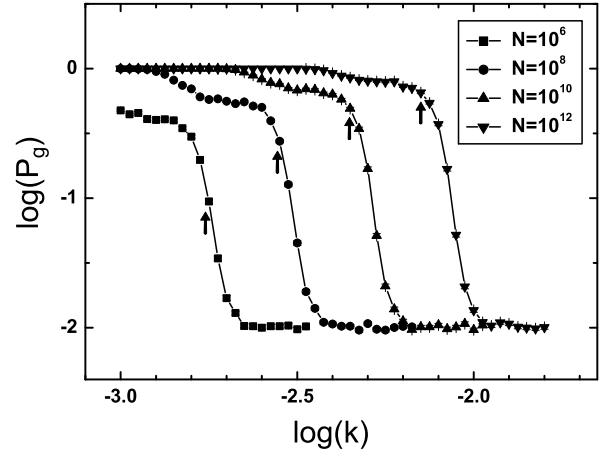}
\caption[0]{\label{fig-5} Logarithm of the growing probability, $\log(P_g)$, as a function of the logarithm of the fitness parameter, $\log(k)$, for $L=10$, $\mu =10^{-3}$, $C=10^{-2}$ and $s=10^{-4}$.
The simulation results for the growing probability in the decoupled continuous-time mutation-selection model are represented by the solid lines with symbols, and the critical fitness parameters, $k_c$, are indicated by the arrows, which correspond, left to right, to the population sizes $N=10^6$, $10^8$, $10^{10}$ and $10^{12}$.}
\end{figure}

Figure \ref{fig-5} shows the logarithm of the growing probability, $\log(P_g)$, as a function of the logarithm of the fitness parameter, $\log(k)$, for $L=10$, $\mu =10^{-3}$, $C=10^{-2}$ and $s=10^{-4}$.
In Fig. \ref{fig-5}, the simulation results for the growing probability in the decoupled continuous-time mutation-selection model are represented by the solid lines with the symbols, which correspond, left to right, to the population sizes $N=10^6$, $10^8$, $10^{10}$ and $10^{12}$.
The growing probability, $P_g$, in Fig. \ref{fig-5} represents the saturated growing probability, $P^*_g$, because $C\gg 1/Ns^*$ for all parameters considered, where $s^* \approx 100 s \approx 10^{-2}$.
The arrows in Fig. \ref{fig-5} indicate the critical fitness parameters, $k_c$, left to right, for $N=10^6$, $10^8$, $10^{10}$ and $10^{12}$, which represent the fitness parameters satisfying $N\Delta X_L^\infty (t=0)=1$.
Figure \ref{fig-5} shows that the saturated growing probability in the deterministic region, $k\ll k_c$, approaches unity and that the saturated growing probability in the stochastic region, $k\gg k_c$, becomes approximately the effective selective advantage, regardless of the fitness parameter and the population size.
The simulation results in Figs. \ref{fig-2}, \ref{fig-3}, \ref{fig-4} and \ref{fig-5} indicate that the selective advantage ratio, $r$, maintains a similar value for various population sizes, selective advantages, measuring parameters, and fitness parameters, if the sequence length $L$ is the same.

\begin{figure}[h]
\includegraphics[width=10cm]{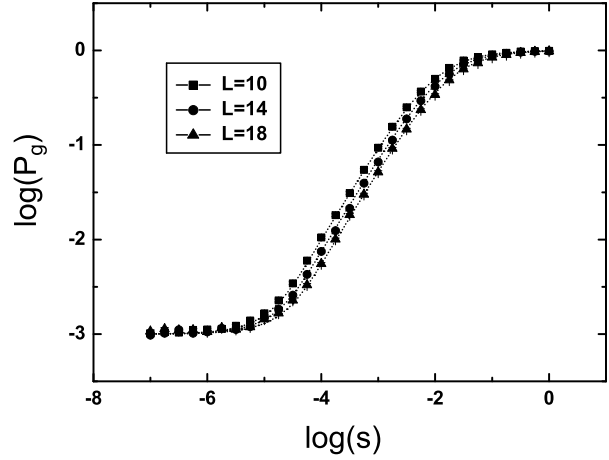}
\caption[0]{\label{fig-6} Logarithm of the growing probability, $\log(P_g)$, as a function of the logarithm of the selective advantage, $\log(s)$, for $\mu =0.001$ and $k=0.002$.
The population sizes are set to $N=10^5$, $10^8$ and $10^{11}$, and the measuring parameters are set to $C=10^{-2}$, $10^{-5}$ and $10^{-8}$, for $L=10$, $14$ and $18$, respectively.
The simulation results for the growing probability in the decoupled continuous-time mutation-selection model are represented by the symbols, which correspond, top to bottom, to the sequence lengths, $L=10$, $14$ and $18$.
The theoretical formula in Eq. (\ref{eq-6}) with $NC=10^3$ and $s^*=rs$ is represented by the short-dotted lines, which correspond, top to bottom, to the selective advantage ratios, $r=100$, 75 and 55.}
\end{figure}

Figure \ref{fig-6} shows the logarithm of the growing probability, $\log(P_g)$, as a function of the logarithm of the selective advantage, $\log(s)$, for $\mu =0.001$ and $k=0.002$. In Fig. \ref{fig-6}, the simulation results for the growing probability in the decoupled continuous-time mutation-selection model are represented by the symbols, that correspond, top to bottom, to the sequence lengths, $L=10$, $14$ and $18$.
The computer simulation for an infinite population showed that the critical population sizes $N_c$ satisfying $N_c\Delta X^\infty_L(t=0)=1$ were approximately $3.98\times 10^6$, $2.48\times 10^9$ and $1.61\times 10^{12}$ for $L=10$, $14$ and $18$, respectively.
Hence, the population sizes were set to $N=10^5$, $10^8$ and $10^{11}$ for $L=10$, $14$ and $18$, respectively, to belong to the stochastic region, $N\ll N_c$.
The measuring parameters were set to $C=10^{-2}$, $10^{-5}$ and $10^{-8}$ for $L=10$, $14$ and $18$, respectively, to maintain the same product of the population size and the measuring parameter, or $NC=10^3$, because the growing probability depends on $NC$, as shown in Eq. (\ref{eq-6}).

Figure \ref{fig-6} shows that the simulation results for the growing probability for $L=10$, $14$ and $18$ have a similar dependence on the selective advantage. 
In Fig. \ref{fig-6}, the theoretical formula in Eq. (\ref{eq-6}) with $NC=10^3$ and $s^*=rs$ is represented by the short-dotted lines, which correspond, top to bottom, to the selective advantage ratios, $r=100$, 75 and 55.
A comparison of the symbols and the short-dotted lines showed that the simulation results for the growing probability, $P_g$, could be fitted well using the theoretical formula for the growing probability in the Moran two-allele model with the adjusted selective advantage ratio, which decreases with  increasing sequence length.
The simulation showed that the selective advantage ratio does not depend on the population size, selective advantage, measuring parameter, and fitness parameter; instead, that the selective advantage ratio decreases with  increasing sequence length.

\section{Summary} \label{sec-4}

The growing probabilities of additional offspring with the beneficial reversal allele for growing to a size, $NC$, for various population sizes, $N$, sequence lengths, $L$, selective advantages, $s$, fitness parameters, $k$, and measuring parameters, $C$, were calculated for a haploid, asexual population in the decoupled continuous-time mutation-selection model using the logistic branching process.
The growing probability was measured from the initial steady state in a sharply-peaked landscape, by switching on the asymmetric sharply-peaked landscape with a positive selective advantage of the reversal allele over the optimal allele. The growing probability was evaluated as the ratio of the first arrival time to the final arrival time, where the first and the final arrival times are defined as the arrival time of the first additional offspring and that of a successful offspring with the reversal allele that can grow to a size, $NC$, respectively.

Computer simulations showed that the growing probability in an asymmetric sharply-peaked landscape in the decoupled continuous-time mutation-selection model in the stochastic region satisfying $N\Delta X^\infty_L(t=0) \ll 1$ was inversely proportional to the measuring parameter when $C<1/Ns^*$, bent when $C\approx 1/Ns^*$ and became saturated when $C>1/Ns^*$, where $s^*$ is the effective selective advantage.
This suggests that the number of additional offspring with the reversal allele, $n_L$, in the stochastic region drifts neutrally until $n_L$ reaches the size, $NC$, when $n_L<1/s^*$, and that it grows deterministically when  $n_L>1/s^*$. 
Computer simulations also showed that the saturated growing probability in the stochastic region was approximately the effective selective advantage when $C\gg 1/Ns^*$ and $s^*\ll 1$.

Computer simulations showed that the growing probability in the stochastic region in an asymmetric sharply-peaked landscape in the decoupled continuous-time mutation-selection model for various population sizes, selective advantages and measuring parameters could be fitted well using the theoretical formula for the growing probability in the Moran two-allele model.
A previous study showed that the growing probability in the stochastic region in the coupled discrete-time mutation-selection model could be described using the theoretical formula for the growing probability in the Wright-Fisher two-allele model\cite{18}.
The present study suggests that the growing probability in the stochastic region in an asymmetric sharply-peaked landscape in the decoupled continuous-time mutation-selection model could be described using the theoretical formula for the growing probability in the Moran two-allele model.

Computer simulations showed that the selective advantage ratio, which represents the ratio of the effective selective advantage to the selective advantage, does not depend on the population size, selective advantage, measuring parameter, and fitness parameter; instead, that the selective advantage ratio decreases with increasing sequence length.
The growth behavior of additional offspring with the reversal allele in an asymmetric sharply-peaked landscape in the decoupled continuous-time mutation-selection model was controlled by the effective selective advantage of the reversal allele over the optimal allele despite there being many other alleles with lower fitness.
The present study might improve the understanding of the evolutionary process in the decoupled continuous-time mutation-selection model.

\begin{acknowledgements}
This study was supported by the Research Fund Program of Research Institute for Basic Sciences, Pusan National University, Korea, 2012, Project No. RIBS-PNU-2012-108.

\end{acknowledgements}

\end{document}